\documentclass[sigconf]{acmart}

\AtBeginDocument{%
  \providecommand\BibTeX{{%
    \normalfont B\kern-0.5em{\scshape i\kern-0.25em b}\kern-0.8em\TeX}}}

\copyrightyear{2023}
\acmYear{2023}
\setcopyright{rightsretained}
\acmConference[CIKM '23]{Proceedings of the 32nd ACM International Conference on Information and Knowledge Management}{October 21--25, 2023}{Birmingham, United Kingdom}
\acmBooktitle{Proceedings of the 32nd ACM International Conference on Information and Knowledge Management (CIKM '23), October 21--25, 2023, Birmingham, United Kingdom}\acmDOI{10.1145/3583780.3614949}
\acmISBN{979-8-4007-0124-5/23/10}

\usepackage{cleveref}
\usepackage{subcaption} 
\usepackage{multirow}
\usepackage[color=lightgray]{todonotes}

\newcommand{\xhdr}[1]{\vspace{1.5ex}\noindent\textbf{#1}}

\newcommand{ \se}[1]{\textcolor{gray}{\footnotesize #1}}

\begin{document}

\title{Large Language Models as Zero-Shot Conversational Recommenders}

\author{Zhankui He}
\authornote{Both authors contributed equally to this research.}
\email{zhh004@eng.ucsd.edu}
\affiliation{
  \institution{University of California, San Diego}
  \city{La Jolla}
  \state{California}
  \country{USA}
}

\author{Zhouhang Xie}
\authornotemark[1]
\email{zhx022@ucsd.edu}
\affiliation{
  \institution{University of California, San Diego}
  \city{La Jolla}
  \state{California}
  \country{USA}
}

\author{Rahul Jha}
\email{rahuljha@netflix.com}
\affiliation{
  \institution{Netflix Inc.}
  \city{Los Gatos}
  \state{California}
  \country{USA}
}

\author{Harald Steck}
\email{hsteck@netflix.com}
\affiliation{
  \institution{Netflix Inc.}
  \city{Los Gatos}
  \state{California}
  \country{USA}
}

\author{Dawen Liang}
\email{dliang@netflix.com}
\affiliation{
  \institution{Netflix Inc.}
  \city{Los Gatos}
  \state{California}
  \country{USA}
}

\author{Yesu Feng}
\email{yfeng@netflix.com}
\affiliation{
  \institution{Netflix Inc.}
  \city{Los Gatos}
  \state{California}
  \country{USA}
}

\author{Bodhisattwa Prasad Majumder}
\email{bmajumde@eng.ucsd.edu}
\affiliation{
  \institution{University of California, San Diego}
  \city{La Jolla}
  \state{California}
  \country{USA}
}

\author{Nathan Kallus}
\email{nkallus@netflix.com}
\affiliation{
  \institution{Netflix Inc.}
  \city{Los Gatos}
  \state{California}
  \country{USA}
}
\affiliation{
  \institution{Cornell University}
  \city{New York}
  \state{New York}
  \country{USA}
}

\author{Julian McAuley}
\email{jmcauley@ucsd.edu}
\affiliation{
  \institution{University of California, San Diego}
  \city{La Jolla}
  \state{California}
  \country{USA}
}

\renewcommand{\shortauthors}{He, et al.}

\begin{abstract}
In this paper, we present empirical studies on conversational recommendation tasks using representative large language models in a zero-shot setting with three primary contributions. \textbf{(1)~Data:} To gain insights into model behavior in ``in-the-wild'' conversational recommendation scenarios, we construct a new dataset of recommendation-related conversations by scraping a popular discussion website. This is the largest public real-world conversational recommendation dataset to date.
\textbf{(2)~Evaluation:} On the new dataset and two existing conversational recommendation datasets, we observe that even without fine-tuning, large language models can outperform existing fine-tuned conversational recommendation models.
\textbf{(3)~Analysis:} We propose various probing tasks to investigate the mechanisms behind the remarkable performance of large language models in conversational recommendation. We analyze both the large language models' behaviors and the characteristics of the datasets, providing a holistic understanding of the models' effectiveness, limitations and suggesting directions for the design of future conversational recommenders.
\end{abstract}



\maketitle

\begin{figure*}[htbp]
  \centering
    \includegraphics[width=\textwidth]{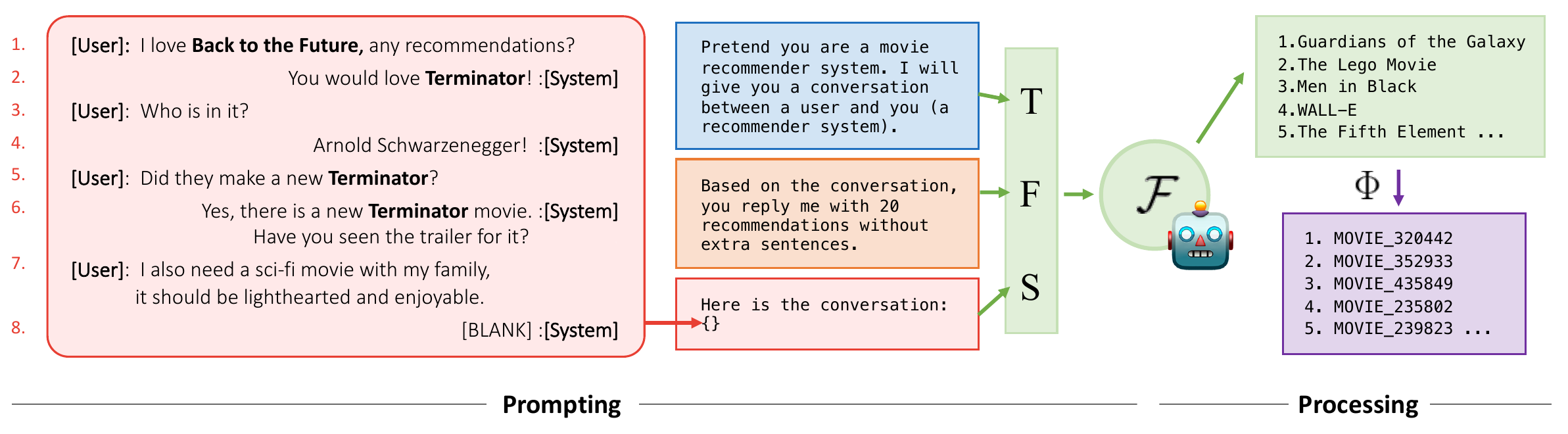}
  \caption{Large Language Models (LLMs) as Zero-Shot Conversational Recommenders (CRS).We introduce a simple prompting strategy to define the task description $T$, format requirement $F$ and conversation context $S$ for a LLM, denoted as $\mathcal{F}$, we then post-process the generative results into ranked item lists with processor $\Phi$.}
  \label{fig:main}
\end{figure*}

 \section{Introduction}
Conversational recommender systems (CRS) aim to elicit user preferences and offer personalized recommendations by engaging in interactive conversations. In contrast to traditional recommenders that primarily rely on users' actions like clicks or purchases, CRS possesses the potential to: (1)~understand not only users' historical actions but also users' (multi-turn) natural-language inputs; (2)~provide not only recommended items but also human-like responses for multiple purposes such as preference refinement, knowledgable discussion or recommendation justification. Towards this objective, a typical  conversational recommender contains two components~\cite{li2018redial, chen2019kbrd, zhou2020kgsf, wang2022unicrs}: a \emph{generator} to generate natural-language responses and a \emph{recommender} to rank items to meet users' needs.

Recently, significant advancements have shown the remarkable potential of large language models (LLMs)\footnote{We refer to LLMs as the large-sized pre-trained language models with exceptional zero-shot abilities as defined in~\cite{zhao2023survey}.}, such as ChatGPT~\cite{openai2022chatgpt}, in various tasks~\cite{brown2020gpt3,openai2023gpt4,zhao2023survey,bubeck2023sparks}. This has captured the attention of the recommender systems community to explore the possibility of leveraging LLMs in recommendation or more general personalization tasks~\cite{kang2023llms,bao2023tallrec,hou2023large,salemi2023lamp,liu2023chatgpt}. Yet,  current efforts generally concentrate on evaluating LLMs in traditional recommendation settings, where only users' past actions like clicks serve as inputs~\cite{kang2023llms,bao2023tallrec,hou2023large,liu2023chatgpt}. The conversational recommendation scenario, though involving more natural language interactions, is still in its infancy~\cite{friedman2023leveraging, wang2023rethinking}.

In this work, we propose to use \emph{large language models as zero-shot conversational recommenders} and then empirically study the LLMs'~\cite{openai2022chatgpt,openai2023gpt4,vicuna2023,xu2023baize} recommendation abilities.
Our detailed contributions in this study include three key aspects regarding \emph{data}, \emph{evaluation}, and \emph{analysis}.

\xhdr{Data.} 
We construct \emph{Reddit-Movie}, a large-scale conversational recommendation dataset with over 634k naturally occurring recommendation seeking dialogs from users from Reddit\footnote{\url{https://www.reddit.com/}}, a popular discussion forum.
Different from existing crowd-sourced conversational recommendation datasets, such as ReDIAL~\cite{li2018redial} and INSPIRED~\cite{hayati2020inspired}, where workers role-play users and recommenders, the \emph{Reddit-Movie} dataset offers a complementary perspective with conversations where users seek and offer item recommendation in the real world.
To the best of our knowledge, this is the largest public conversational recommendation dataset, with 50 times more conversations than ReDIAL.

\xhdr{Evaluation.} By evaluating the recommendation performance of LLMs on multiple CRS datasets, we first notice a \emph{repeated item shortcut} in current CRS evaluation protocols. Specifically, there exist ``repeated items'' in previous evaluation testing samples serving as ground-truth items, which allows the creation of a trivial baseline (e.g.,~copying the mentioned items from the current conversation history) that outperforms most existing models, leading to spurious conclusions regarding current CRS recommendation abilities. After removing the ``repeated items'' in training and testing data, we re-evaluate multiple representative conversational recommendation models~\cite{li2018redial, chen2019kbrd, zhou2020kgsf, wang2022unicrs} on ReDIAL, INSPIRED and our Reddit dataset. With this experimental setup, we empirically show that LLMs can outperform existing fine-tuned conversational recommendation models even without fine-tuning.

\xhdr{Analysis.} 
In light of the impressive performance of LLMs as zero-shot CRS, a fundamental question arises: \emph{What accounts for their remarkable performance?} Similar to the approach taken in~\cite{penha2020prob}, we posit that LLMs leverage both \textit{content/context knowledge} (e.g., ``genre'', ``actors'' and ``mood'') and \textit{collaborative knowledge} (e.g., ``users who like A typically also like B'') to make conversational recommendations. We design several probing tasks to uncover the model's workings and the characteristics of the CRS data. Additionally, we present empirical findings that highlight certain limitations of LLMs as zero-shot CRS, despite their effectiveness.

We summarize the key findings of this paper as follows:

\begin{itemize}
    \item CRS recommendation abilities should be reassessed by eliminating repeated items as ground truth.
    \item LLMs, as zero-shot conversational recommenders, demonstrate improved performance on established and new datasets over fine-tuned CRS models.
    \item LLMs primarily use their superior content/context knowledge, rather than their collaborative knowledge, to make recommendations. 
    \item CRS datasets inherently contain a high level of content/context information, making CRS tasks better-suited for LLMs than traditional recommendation tasks.
    \item LLMs  
    suffer from limitations such as popularity bias and sensitivity to geographical regions.
\end{itemize}

These findings reveal the unique importance of the superior content/context
knowledge in LLMs for CRS tasks, offering great potential to LLMs as an effective approach in CRS; meanwhile,  analyses must recognize the challenges in evaluation, datasets, and potential problems (e.g.,~debiasing) in future CRS design with LLMs.

\section{LLMs as Zero-shot CRS}

\subsection{Task Formation}

Given a user set $\mathcal{U}$, an item set $\mathcal{I}$ and a vocabulary $\mathcal{V}$, a conversation can be denoted as  $C = (u_t, s_t, \mathcal{I}_t)^T_{t=1}$. That means during the $t^\text{th}$ turn of the conversation, a speaker $u_t\in \mathcal{U}$ generates an utterance $s_t=(w_i)^m_{i=1}$, which is a sequence of words $w_i \in \mathcal{V}$. This utterance $s_t$ also contains a set of mentioned items $\mathcal{I}_t\subset \mathcal{I}$ ($\mathcal{I}_t$ can be an empty set if no items mentioned). Typically, there are two users in the conversation $C$ playing the role of \emph{seeker} and \emph{recommender} respectively. Let us use the $2^\text{nd}$ conversation turn in \Cref{fig:main} as an example. Here $t=2$, $u_t$ is \texttt{[System]}, $s_t$ is ``You would love Terminator !'' and $\mathcal{I}_2$ is a set containing the movie \texttt{Terminator}.

Following many CRS papers~\cite{li2018redial, chen2019kbrd, zhou2020kgsf, wang2022unicrs}, the \textit{recommender} component of a CRS is specifically designed to optimize the following objective: during the $k^\text{th}$ turn of a conversation, where $u_k$ is the \emph{recommender}, the recommender takes the conversational context $(u_t, s_t, \mathcal{I}_t)^{k-1}_{t=1}$ as its input, and generate a ranked list of items $\hat{\mathcal{I}}_k$ that best matches the ground-truth items in $\mathcal{I}_k$.

\subsection{Framework}

\xhdr{Prompting.} Our goal is to utilize LLMs as zero-shot conversational recommenders. Specifically, without the need for fine-tuning, we intend to prompt an LLM, denoted as $\mathcal{F}$, using a task description template $T$, format requirement $F$, and conversational context $S$ before the $k^\text{th}$ turn. This process can be formally represented as:
\begin{equation}
\label{eq:LLM}
\hat{\mathcal{I}}_k = \Phi\left(\mathcal{F}(T, F, S)\right).
\end{equation}
To better understand this zero-shot recommender, we present an example in~\Cref{fig:main} with the prompt setup in our experiments.\footnote{We leave more prompting techniques such as CoT~\cite{wei2022chain} in future work.}

\xhdr{Models.} We consider several popular LLMs $\mathcal{F}$ that exhibit zero-shot prompting abilities in two groups. To try to ensure deterministic results, we set the decoding temperature to 0 for all models.
\begin{itemize}
    \item \texttt{\textbf{GPT-3.5-turbo}}~\cite{openai2022chatgpt}\footnote{Referred as \texttt{\textbf{GPT-3.5-t}} hereafter} and \texttt{\textbf{GPT-4}}~\cite{openai2023gpt4} from OPENAI with abilities of solving many complex tasks in zero-shot setting~\cite{openai2023gpt4, bubeck2023sparks} but are closed-sourced.
    \item \textbf{BAIZE}~\cite{xu2023baize}\footnote{We use \textbf{BAIZE-V2} in \url{https://huggingface.co/project-baize/baize-v2-13b}} and \textbf{Vicuna}~\cite{vicuna2023}, which are 
    representative open-sourced LLMs fine-tuned based on \texttt{LLAMA-13B}~\cite{touvron2023llama}.
\end{itemize}

\xhdr{Processing.} We do not assess model weights or output logits from LLMs. Therefore, we apply a post-processor $\Phi$ (e.g.,~fuzzy matching) to convert a recommendation list in natural language to a ranked list $\hat{\mathcal{I}_k}$. The approach of generating item titles instead of ranking item IDs is referred to as a \textit{generative retrieval}~\cite{cao2021autoregressive, tay2022transformer} paradigm. 

\begin{figure}[tbp]
  \centering
    \includegraphics[width=0.95\columnwidth]{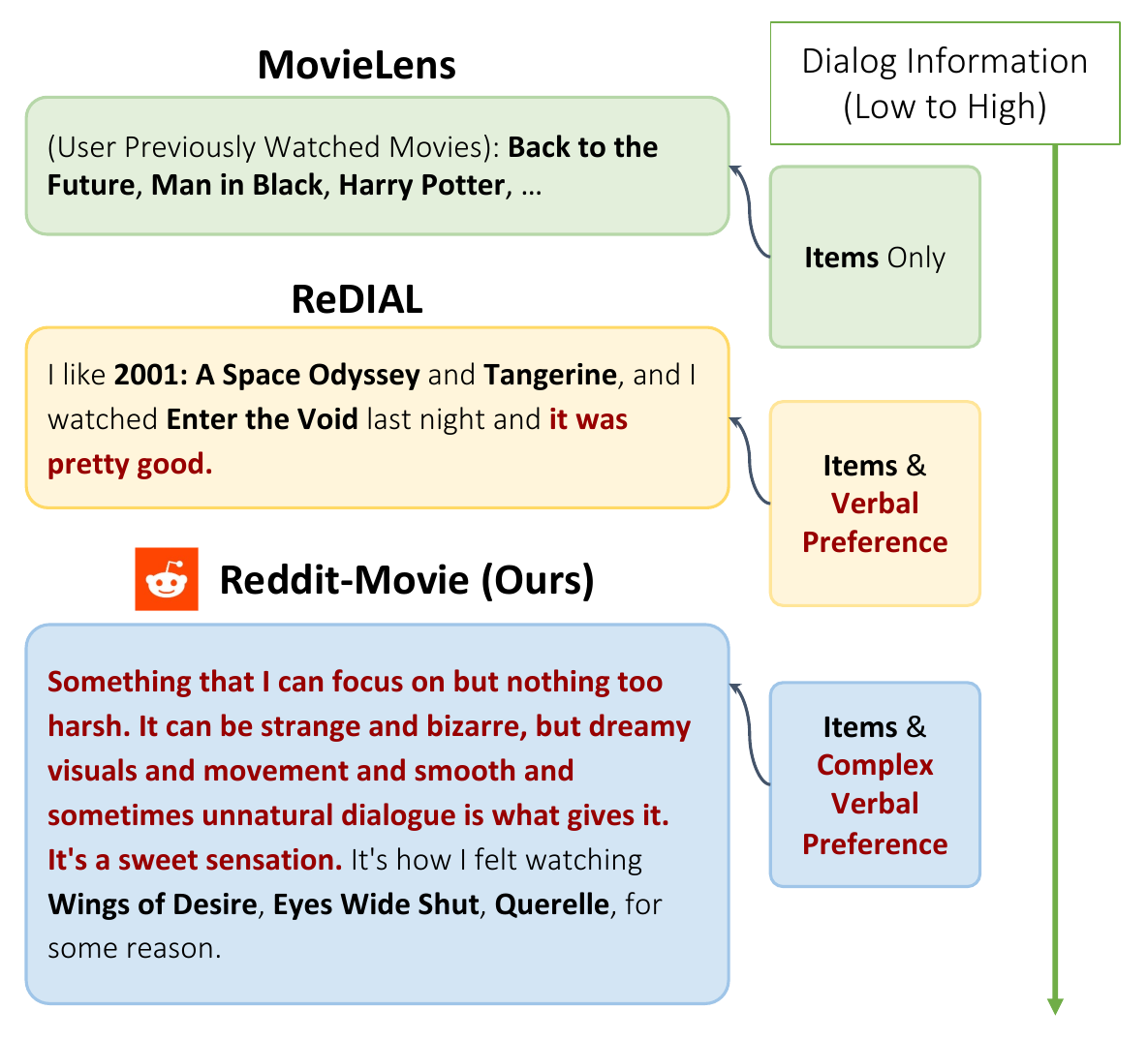}
  \caption{Typical model inputs from a traditional recommendation dataset (MovieLens~\cite{Harper2016TheMD}), an existing CRS dataset (ReDIAL~\cite{li2018redial}), and our \emph{Reddit-Movie} dataset. The \emph{Reddit-Movie} dataset contains more information in its textual content compared to existing datasets where users often explicitly specify their preference. See \Cref{sec:dataset-comparison} for quantitative analysis.}
  \label{fig:dataset-examples}
\end{figure}

\begin{table}[t]
\small
\caption{Dataset Statistics. We denote a subset of \emph{Reddit-Movie} in 2022 as  \texttt{base}, and the entire ten-year dataset as \texttt{large}.}
\label{tab:stats}
\begin{tabular}{lccccc}
\toprule
\textbf{Dataset}                  & \textbf{\#Conv.} & \textbf{\#Turns} & \textbf{\#Users} & \textbf{\#Items}  \\ \midrule
\textbf{INSPIRED}~\cite{hayati2020inspired}                 & 999                      & 35,686           & 999              & 1,967      \\
\textbf{ReDIAL}~\cite{li2018redial}                   & 11,348                 & 139,557          & 764              & 6,281        \\
\textbf{\textit{Reddit-Movie}}$^\texttt{base}$      &    85,052               &   133,005        & 10,946           &    24,326   \\
\textbf{\textit{Reddit-Movie}}$^\texttt{large}$ & 634,392                  & 1,669,720        & 36,247           & 51,203     \\ \bottomrule
\end{tabular}
\vspace{-5pt}
\end{table}

\begin{figure*}[htbp]
  \centering
  \begin{subfigure}[b]{0.48\textwidth}
    \includegraphics[width=\textwidth]{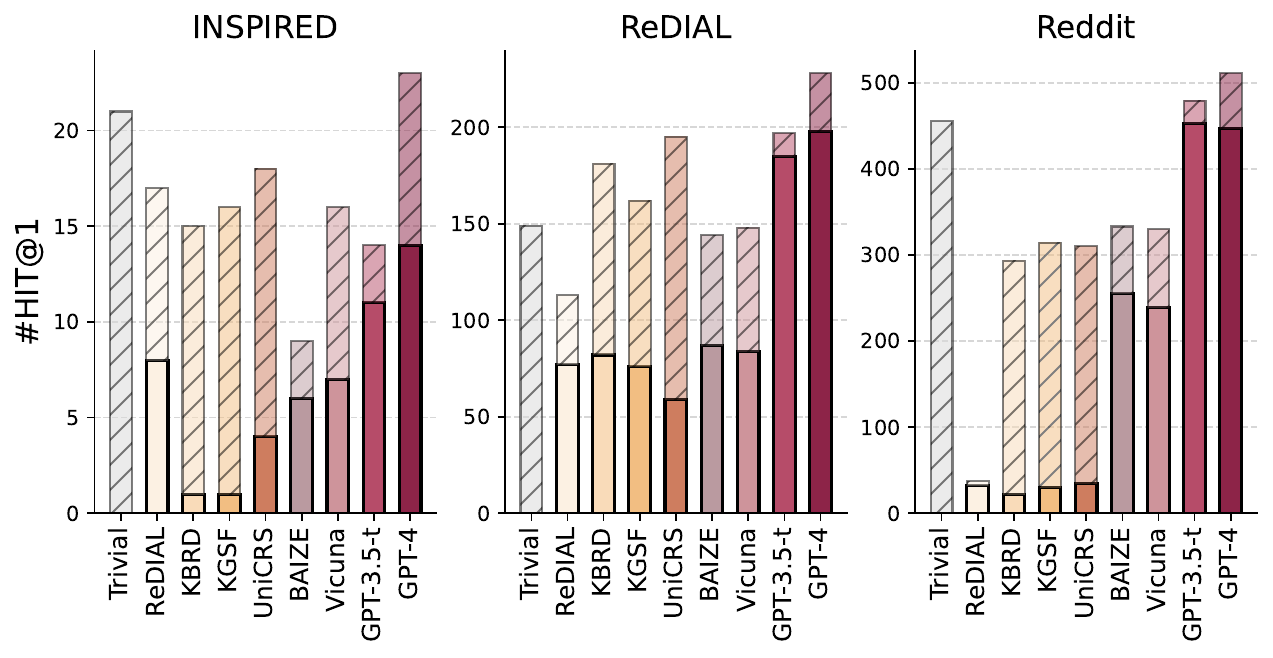}
  \end{subfigure}
  \hfill
  \begin{subfigure}[b]{0.48\textwidth}
    \includegraphics[width=\textwidth]{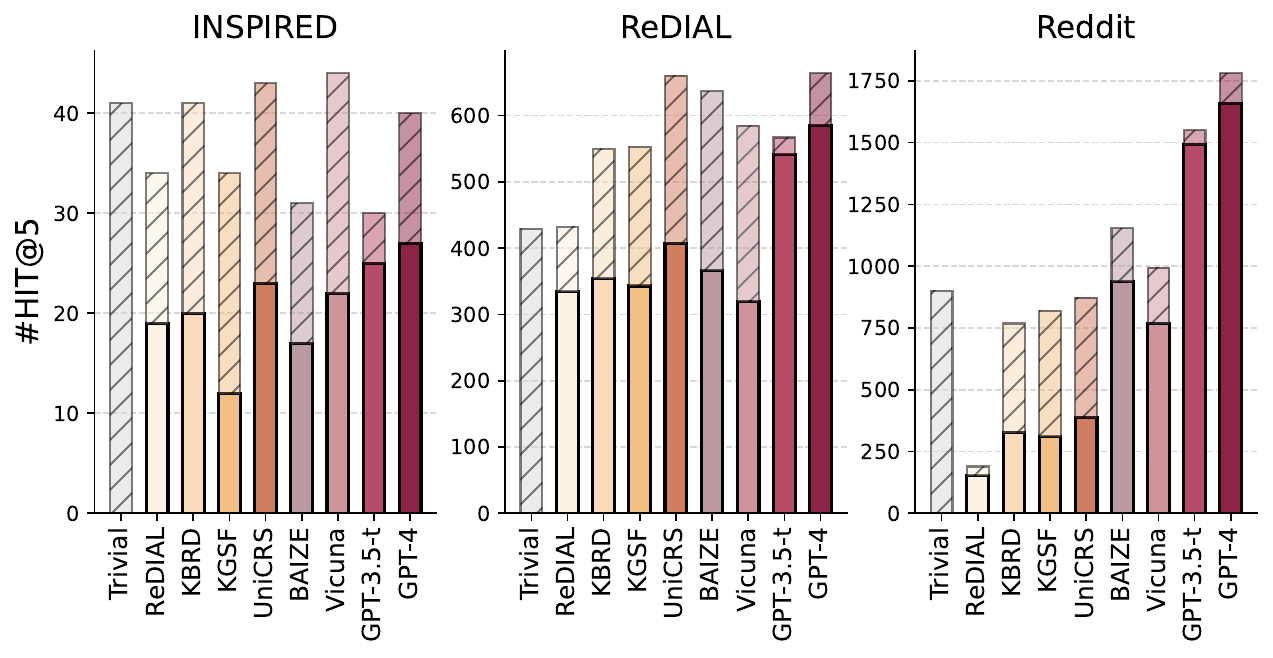}
  \end{subfigure}
  \caption{To show the \textit{repeated item shortcut}, we count CRS recommendation hits using the Top-K ranked list $K=\{1,5\}$. We group the ground-truth hits by repeated items (shaded bars) and new items (not shaded bars). The trivial baseline copies existing items from the current conversation history in chronological order,  from the most recent and does not recommend new items. }
  \label{fig:repetition}
\end{figure*}

\section{Dataset}
\label{sec:dataset}

Ideally, a large-scale dataset with diverse interactions and real-world conversations is needed to evaluate models' ability in conversational recommendation. Existing conversational recommendation datasets are usually crowd-sourced~\cite{li2018redial, hayati2020inspired, kang2019self-play, zhou2020towards} and thus only partially capture realistic conversation dynamics. For example, a crowd worker responded with \texttt{"Whatever Whatever I'm open to any suggestion."} when asked about movie preferences in ReDIAL;
this happens since crowd workers often do not have a particular preference at the time of completing a task. In contrast, a real user could have a very particular need, as shown in \Cref{fig:dataset-examples}.

To complement crowd-sourced CRS datasets, we present the \emph{Reddit-Movie} dataset, the largest-scale conversational movie recommendation dataset to date, with naturally occurring movie recommendation conversations that can be used along with existing crowd-sourced datasets to provide richer perspectives for training and evaluating CRS models. In this work, we conduct our model evaluation and analysis on two commonly used crowd-sourcing datasets: ReDIAL~\cite{li2018redial} and INSPIRED~\cite{hayati2020inspired}, as well as our newly collected Reddit dataset. We show qualitative examples from the Reddit dataset as in~\Cref{fig:dataset-examples} and quantitative analysis in \Cref{sec:dataset-comparison}.

\xhdr{Dataset Construction} 
To construct a CRS dataset from Reddit, we process all Reddit posts from 2012 Jan to 2022 Dec from \emph{pushshift.io}\footnote{\url{https://pushshift.io/}}. 
We consider movie recommendation scenarios\footnote{Other domains like songs, books can potentially be processed in a similar way} and extract related posts from five related subreddits: \emph{r/movies}, \emph{r/bestofnetflix}, \emph{r/moviesuggestions}, \emph{r/netflixbestof} and \emph{r/truefilm}. We process the raw data with the pipeline of \emph{conversational recommendation identification}, \emph{movie mention recognition} and \emph{movie entity linking}\footnote{Check our evaluation data, LLMs scripts, results and the links of Reddit-Movie datasets in \url{https://github.com/AaronHeee/LLMs-as-Zero-Shot-Conversational-RecSys}.}. In our following evaluation, we use the most recent 9k conversations in \emph{Reddit-Movie}$^\texttt{base}$ from December 2022 as the testing set since these samples occur \textit{after} GPT-3.5-t's release. Meanwhile, GPT-4~\cite{openai2023gpt4} also mentioned its pre-training data cut off in Sept. 2021\footnote{We note that there is a possibility that GPT-4's newest checkpoint might include a small amount of more recent data~\cite{openai2023gpt4}.}. For other compared models, we use the remaining 76k conversations in \emph{Reddit-Movie}$^\texttt{base}$ dataset for training and validation.

\xhdr{Discussion.} 
From the statistics in~\Cref{tab:stats}, we observe: (1) The dataset \emph{Reddit-Movie} stands out as the largest conversational recommendation dataset, encompassing 634,392 conversations and covering 51,203 movies. (2) In comparison to ReDIAL~\cite{li2018redial} and INSPIRED~\cite{hayati2020inspired}, \emph{Reddit-Movie} contains fewer multi-turn conversations, mainly due to the inherent characteristics of Reddit posts. (3) By examining representative examples depicted in~\Cref{fig:dataset-examples}, we find that \emph{Reddit-Movie} conversations tend to include more complex and detailed user preference in contrast to ReDIAL, as they originate from real-world conversations on Reddit, enriching the conversational recommendation datasets with a diverse range of discussions. 

\begin{figure*}[htbp]
  \centering
  \begin{subfigure}[b]{0.48\textwidth}
    \includegraphics[width=\textwidth]{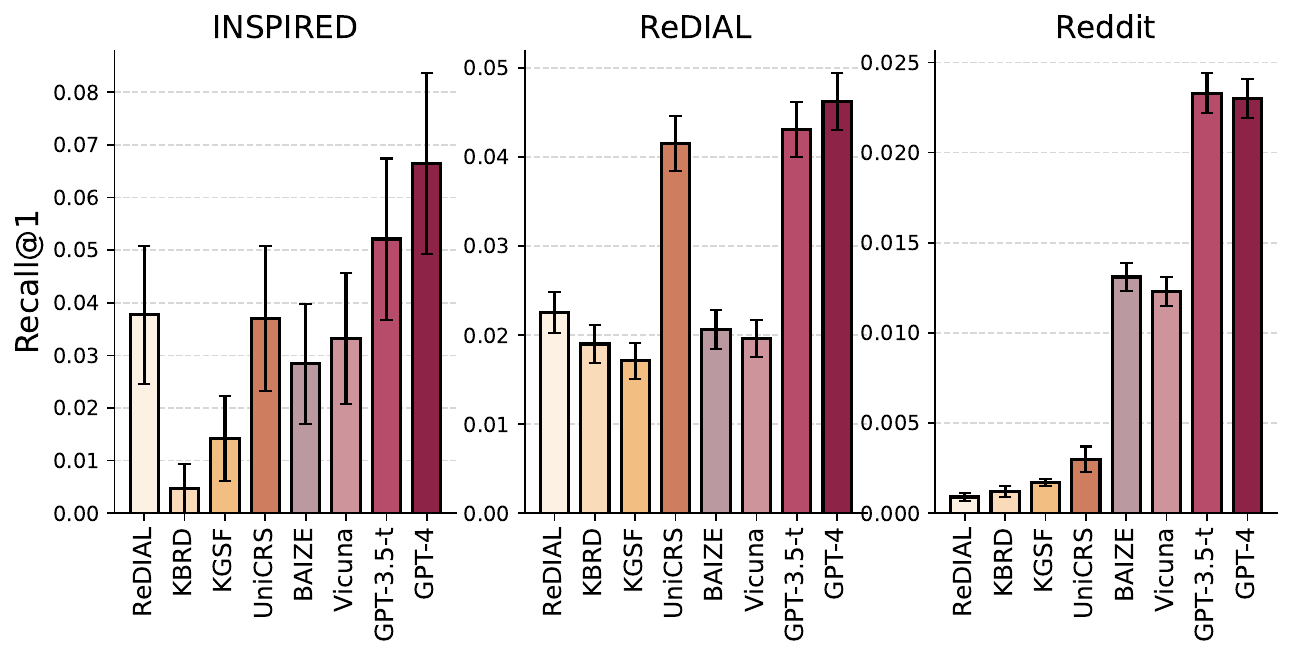}
  \end{subfigure}
  \hfill
  \begin{subfigure}[b]{0.48\textwidth}
    \includegraphics[width=\textwidth]{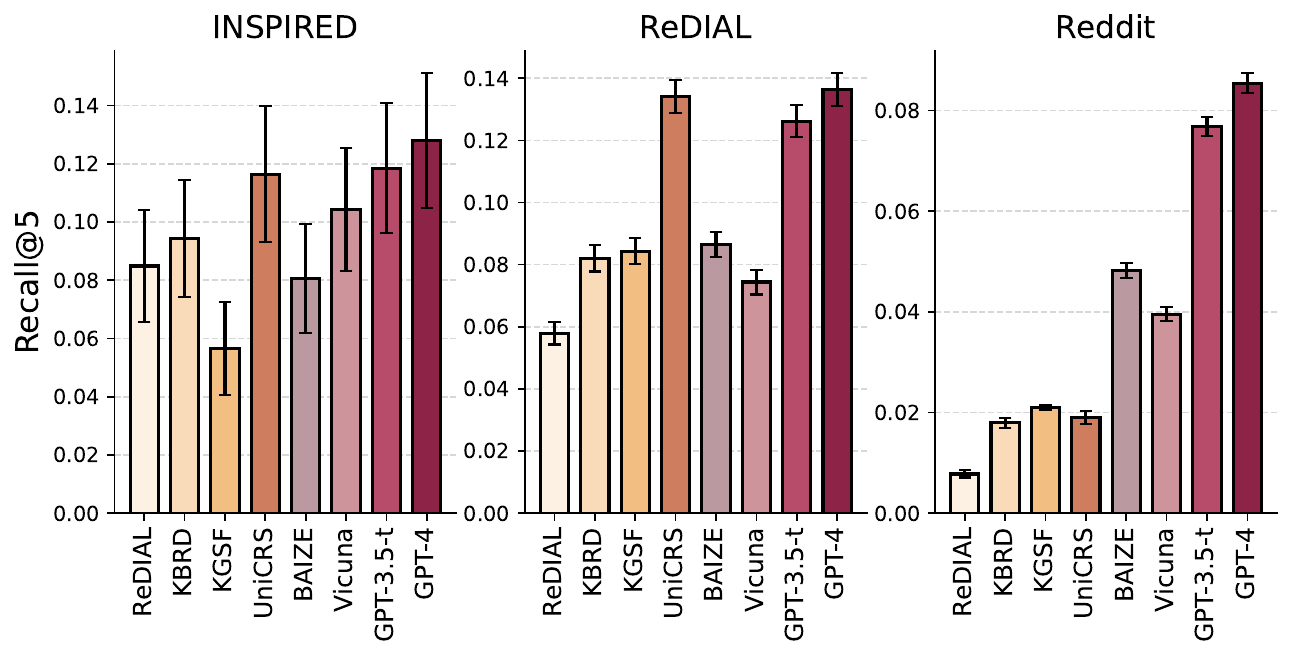}
  \end{subfigure}
  \caption{CRS recommendation performance on New Items in terms of Recall@K, with $K=\{1,5\}$. To exclude the influence of repeated items in CRS evaluation, we remove all repeated items in training and testing datasets and re-train all baselines.}
  \label{fig:new}
\end{figure*}

\section{Evaluation}

In this section, we evaluate the proposed LLMs-based frameowrk on ReDIAL~\cite{li2018redial}, INSPIRED~\cite{hayati2020inspired} and our Reddit datasets. We first explain the evaluation setup and a \textit{repeated item shortcut} of the previous evaluation in~\Cref{sec:eval_setup,sec:repeat}. Then, we re-train models and discuss LLM performance in~\Cref{sec:general_discussion}.

\subsection{Evaluation Setup}
\label{sec:eval_setup}

\xhdr{Repeated \textit{vs.}~New Items.} 
Given a conversation $C=(u_t, s_t, \mathcal{I}_t)^T_{t=1}$, it is challenging to identify the ground-truth recommended items, i.e.,~whether the mentioned items $\mathcal{I}_k$ at the $k^\text{th} (k\le T)$ turn are used for recommendation purposes. A common evaluation setup assumes that when $u_k$ is the \emph{recommender}, all items $i \in \mathcal{I}_k$ serve as ground-truth recommended items.

In this work, we further split the items $i\in\mathcal{I}_k$ into two categories: \textit{repeated items} or \textit{new items}. Repeated items are items that have appeared in previous conversation turns, i.e., $\{i \mid \exists t \in [1, k), i \in \mathcal{I}_{t}\}$; and new items are items not mentioned in previous conversation turns. We explain the details of this categorization in~\Cref{sec:repeat}.

\xhdr{Evaluation Protocol.} On those three datasets, we evaluate several representative CRS models and several LLMs on their recommendation abilities. For baselines, after re-running the training code provided by the authors, we report the prediction performance using Recall@K~\cite{li2018redial, chen2019kbrd, zhou2020kgsf, wang2022unicrs} (i.e.,~HIT@K). We consider the means and the standard errors\footnote{We show standard errors as error bars in our figures and gray numbers in our tables.} of the metric with $K=\{1,5\}$.

\xhdr{Compared CRS Models.} We consider several representative CRS models. For baselines which rely on structured knowledge, we use the entity linking results of ReDIAL and INSPIRED datasets provided by UniCRS~\cite{wang2022unicrs}. Note that we do not include more works~\cite{li2022uccr,ren2022upcr,ma2021crwalker} because UniCRS~\cite{wang2022unicrs} is representative with similar results.

\begin{itemize}
    \item \textbf{ReDIAL}~\cite{li2018redial}: This model is released along with the ReDIAL dataset with an auto-encoder~\cite{sedhain2015autorec}-based recommender.
    \item \textbf{KBRD}~\cite{chen2019kbrd}: This model proposes to use the DBPedia~\cite{auer2007dbpedia} to enhance the semantic knowledge of items or entities. 
    \item \textbf{KGSF}~\cite{zhou2020kgsf}: This model incorporates two knowledge graphs to enhance the representations of words and entities, and uses the Mutual Information Maximization method to align the semantic spaces of those two knowledge graphs. 
    \item \textbf{UniCRS}~\cite{wang2022unicrs}: This model uses pre-trained language model, DialoGPT~\cite{zhang2020dialogpt}, with prompt tuning to conduct recommendation and conversation generation tasks respectively. 
\end{itemize}

\subsection{Repeated Items Can Be Shortcuts}
\label{sec:repeat}

Current evaluation for conversational recommendation systems does not differentiate between repeated and new items in a conversation. We observed that this evaluation scheme favors systems that optimize for mentioning repeated items. As shown in~\Cref{fig:repetition}, a trivial baseline that always copies seen items from the conversation history has better performance than most previous models under the standard evaluation scheme.
This phenomenon highlights the risk of shortcut learning~\cite{Geirhos2020ShortcutLI}, where a decision rule performs well against certain benchmarks and evaluations but fails to capture the true intent of the system designer. Indeed, the \#HIT@1 for the models tested dropped by more than 60\% on average when we focus on new item recommendation only, which is unclear from the overall recommendation performance. After manually checking, we observe a typical pattern of repeated items, which is shown in the example conversation in ~\Cref{fig:main}. In this conversation, \texttt{Terminator} at the $6^\text{th}$ turn is used as the ground-truth item. The system repeated this \texttt{Terminator} because the system quoted this movie for a content-based discussion during the conversation rather than making recommendations. Given the nature of recommendation conversations between two users, it is more probable that items repeated during a conversation are intended for discussion rather than serving as recommendations. We argue that considering the large portion of repeated items (e.g.,~more than 15\% ground-truth items are repeated items in INSPIRED), it is beneficial to remove repeated items and re-evaluate CRS models to better understand models' recommendation ability. It is worth noting that the repetition patterns have also been investigated in evaluating other recommender systems such as \textit{next-basket} recommendation~\cite{li2023next}.

\begin{figure*}[htbp]
  \centering
    \includegraphics[width=0.9\textwidth]{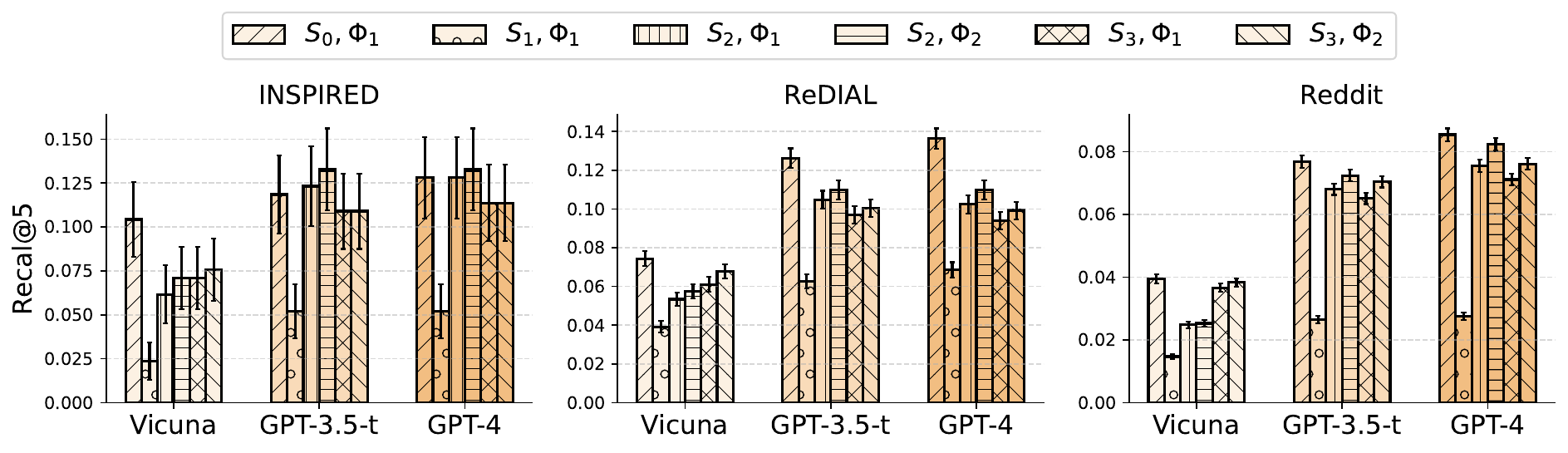}
    \caption{Ablation studies for the research question about the primary knowledge used by LLMs for CRS. Here $\Phi_1$ is the post-processor which only considers in-dataset item titles; $\Phi_2$ is the post-processor based on $\Phi_1$ and further excludes all seen items in conversational context from generated recommendation lists. For inputs like \textit{Original} ($S_0$) and \textit{ItemOnly} ($S_1$), LLMs show similar performance with $\Phi_1$ or $\Phi_2$, so we only keep $\Phi_1$ here. We consider $\Phi_2$ because \textit{ItemRemoved}~($S_2$) and \textit{ItemRandom}~($S_3$) have no information about already mentioned items, which may cause under-estimated accuracy using $\Phi_1$ compared to \textit{Original}.}
  \label{fig:primary}
\end{figure*}

\subsection{LLMs Performance}
\label{sec:general_discussion}

\xhdr{Finding 1 - LLMs outperform fine-tuned CRS models in a zero-shot setting.}
For a comparison between models' abilities to recommend new items to the user in conversation, we re-train existing CRS models on all datasets for new item recommendation only. The evaluation results are as shown in~\Cref{fig:new}. Large language models, although not fine-tuned, have the best performance on all datasets.
Meanwhile, the performance of all models is uniformly lower on Reddit compared to the other datasets, potentially due to the large number of items and fewer conversation turns, making recommendation more challenging.

\begin{table}[]
\small
\caption{Recall@1 results of considering all generated item titles ($\Phi_0$) and only considering in-dataset item titles ($\Phi_1$).}
\resizebox{\columnwidth}{!}{
\begin{tabular}{lcccccc}
\toprule
              & \multicolumn{2}{c}{\textbf{INSPIRED}}                & \multicolumn{2}{c}{\textbf{ReDIAL}}                  & \multicolumn{2}{c}{\textbf{Reddit}}                  \\ \cmidrule(l){2-3} \cmidrule(l){4-5} \cmidrule(l){6-7} 
\textbf{Model}   & $\Phi_0$             & $\Phi_1$             & $\Phi_0$             & $\Phi_1$             & $\Phi_0$             & $\Phi_1$             \\ \midrule
\textbf{BAIZE}         &   .019 \se{.019}   &     \textbf{.028} \se{.011}                 & \textbf{.021} \se{.002}   &  \textbf{.021} \se{.002}       &    .012 \se{.001}       &    \textbf{.013} \se{.008}         \\
\textbf{Vicuna}        & .028 \se{.011}         &  \textbf{.033} \se{.012}                    &   \textbf{.020} \se{.002}                   &  \textbf{.020} \se{.002}                   &    \textbf{.012} \se{.001}                  & \textbf{.012} \se{.001}                  \\
\textbf{\texttt{GPT-3.5-t}} &  .047 \se{.015}  & \textbf{.052} \se{.015} & .041 \se{.003} & \textbf{.043} \se{.003} & .022 \se{.001} & \textbf{.023} \se{.001} \\
\textbf{\texttt{GPT-4}}         &  .062 \se{.017}   &  \textbf{.066} \se{.017}   & .043 \se{.003}   &   \textbf{.046} \se{.004}  &  .022 \se{.001}    &    \textbf{.023} \se{.001}  \\ \bottomrule
\end{tabular}
}
\label{tab:phi}
\end{table}

\xhdr{Finding 2 - GPT-based models achieve superior performance than  open-sourced LLMs.} 
As shown in \Cref{fig:new}, large language models consistently outperform other models across all three datasets, while GPT-4 is generally better than GPT-3.5-t. We hypothesize this is due to GPT-4's larger parameter size enables it to retain more correlation information between movie names and user preferences that naturally occurs in the language models' pre-training data. Vicuna and BAIZE, while having comparable performance to prior models on most datasets, have significantly lower performance than its teacher, GPT-3.5-t. This is consistent with previous works' finding that smaller distilled models via imitation learning cannot fully inherit larger models ability on downstream tasks~\cite{gudibande2023false}.

\begin{table}[]
\caption{Fraction of Top-K ($K=20$ in our prompt setup) recommendations (\#rec) that can be string matched in the IMDB movie database (\%imdb) for the different models, which shows a lower bound of non-hallucinated movie titles.}
\resizebox{\columnwidth}{!}{%
\begin{tabular}{cccccccc}
\toprule
\multicolumn{2}{c}{\textbf{BAIZE}} & \multicolumn{2}{c}{\textbf{Vicuna}} & \multicolumn{2}{c}{\textbf{\texttt{GPT-3.5-t}}} & \multicolumn{2}{c}{\textbf{\texttt{GPT-4}}} \\ \cmidrule(l){1-2} \cmidrule(l){3-4} \cmidrule(l){5-6}  \cmidrule(l){7-8} 
\#rec            & \%imdb         & \#rec            & \%imdb           & \#rec              & \%imdb            & \#rec            & \%imdb          \\ \midrule
259,333          & 81.56\%      &  258,984          & 86.98\%          & 321,048            & 95.51\%           & 322,323          & 94.86\%         \\ \bottomrule
\label{tab:ha}
\end{tabular}
}
\vspace{-10pt}
\end{table}

\xhdr{Finding 3 - LLMs may generate out-of-dataset item titles, but few hallucinated recommendations.}
We note that language models trained on open-domain data naturally produce items out of the allowed item set during generation. In practice, removing these items improves the models' recommendation performance. Large language models outperform other models (with GPT-4 being the best) consistently regardless of whether these unknown items are removed or not, as shown in~\Cref{tab:phi}. Meanwhile, \Cref{tab:ha} shows that around 95\% generated recommendations from GPT-based models (around 81\% from BAIZE and 87\% from Vicuna) can be found in IMDB~\footnote{Movie titles in \url{https://datasets.imdbws.com/}.} by string matching. Those lower bounds of these matching rates indicate that there are only a few hallucinated item titles in the LLM recommendations in the movie domain. 
 
\section{Detailed Analysis}

Observing LLMs' remarkable conversational recommendation performance for zero-shot recommendation, we are interested in \textit{what accounts for their
effectiveness} and \textit{what their limitations are}. We aim to answer these questions from both a model and data perspective. 

\begin{figure}[htbp]
  \centering
    \includegraphics[width=\columnwidth]{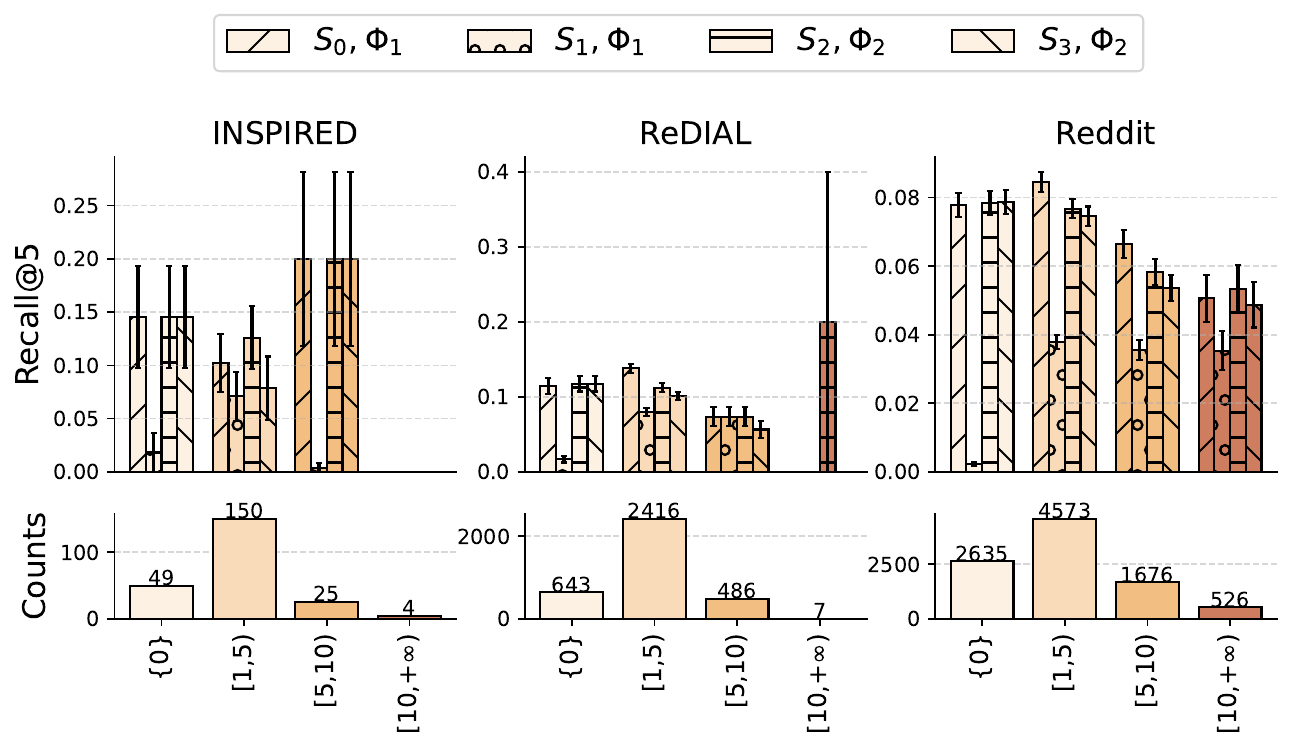}
  \caption{\texttt{GPT-3.5-t} Recall@5 results grouped by the occurrences of items in conversation context, and count the conversations per dataset.}
  \label{fig:occurence}
\end{figure}

\subsection{Knowledge in LLMs}
\label{sec: know_llm}

\xhdr{Experiment Setup.} Motivated by the probing work of~\cite{penha2020prob}, we posit that two types of knowledge in LLMs can be used in CRS:

\begin{itemize}
    \item \textbf{Collaborative knowledge}, which requires the model to match items with similar ones, according to community interactions like ``users who like A typically also like B''. In our experiments, we define the collaborative knowledge in LLMs as \textit{the ability to make accurate recommendations using item mentions in conversational contexts}.
    \item \textbf{Content/context knowledge}, which requires the model to match recommended items with their content or context information.  In our experiments, we define the content/context knowledge in LLMs as \textit{the ability to make accurate recommendations based on all other conversation inputs rather than item mentions,} such as contextual descriptions, mentioned genres, and director names.
    
\end{itemize}

To understand how  LLMs use these two types of knowledge, given the original conversation context $S$ (Example in~\Cref{fig:main}), we perturb $S$ with three different strategies as follows and subsequently re-query the LLMs. We denote the original as $S_0$:

\begin{itemize}
    \item \textbf{$\mathbf{S_0}$ (Original):} we use the original conversation context.
    \item \textbf{$\mathbf{S_1}$ (ItemOnly):} we keep mentioned items and remove all natural language descriptions in the conversation context. 
    \item \textbf{$\mathbf{S_2}$ (ItemRemoved):} we remove mentioned items and keep other content in the conversation context. 
    \item \textbf{$\mathbf{S_3}$ (ItemRandom):} we replace the mentioned items in the conversation context with items that are uniformly sampled from the item set $\mathcal{I}$ of this dataset, to eliminate the potential influence of $S_2$ on the sentence grammar structure.
\end{itemize}

\begin{table}[t]
\small
\caption{To understand the content/context knowledge in LLMs and existing CRS models, we re-train the existing CRS models using the same perturbed conversation context \textit{ItemRemoved} ($S_2$). We include the results of the representative CRS model UniCRS (denoted as CRS*) as well as a representative text-encoder \texttt{BERT-small}~\cite{devlin2019bert} (denoted as TextEnc*).}
\resizebox{\columnwidth}{!}{
\begin{tabular}{lcccccc}
\toprule
                    & \multicolumn{2}{c}{\textbf{INSPIRED}}       & \multicolumn{2}{c}{\textbf{ReDIAL}}         & \multicolumn{2}{c}{\textbf{Reddit}}         \\ \cmidrule(l){2-3} \cmidrule(l){4-5} \cmidrule(l){6-7}
\textbf{Model}     & R@1                  & R@5                  & R@1                  & R@5                  & R@1                  & R@5                  \\ \midrule
\textbf{Vicuna}             & .024 \se{.010} & .062 \se{.017} & .014 \se{.002} & .053 \se{.003} & .008 \se{.001} & .025 \se{.001} \\
\textbf{\texttt{GPT-3.5-t}} & .057 \se{.016} & .123 \se{.023} & .030 \se{.003} & \textbf{.105} \se{.005} & .018 \se{.001} & .068 \se{.002} \\
\textbf{\texttt{GPT-4}}     & \textbf{.062} \se{.017} & \textbf{.128} \se{.023} & \textbf{.032} \se{.003} & .102 \se{.005} & \textbf{.019} \se{.001} & \textbf{.075} \se{.002} \\ \midrule
\textbf{CRS*}               & .039 \se{.011} & .087 \se{.014} & .015 \se{.002} & .058 \se{.003} & .001 \se{.000} & .008 \se{.001} \\
\textbf{TextEnc*}           & .038 \se{.015} & .090 \se{.016} & .013 \se{.002} & .053 \se{.004} & .002 \se{.000} & .009 \se{.001} \\ \bottomrule
\end{tabular}
}
\label{tab:content_know}
\end{table}

\xhdr{Finding 4 - LLMs mainly rely on content/context knowledge to make recommendations.} \Cref{fig:primary} shows a drop in performance for most models across various datasets when replacing the original conversation text \textit{Original} ($S_0$) with other texts, indicating that LLMs leverage both content/context knowledge and collaborative knowledge in recommendation tasks. However, the importance of these knowledge types differs. Our analysis reveals that content/context knowledge is the primary knowledge utilized by LLMs in CRS. When using \textit{ItemOnly} ($S_1$) as a replacement for \textit{Original}, there is an average performance drop of more than 60\% in terms of Recall@5. On the other hand, GPT-based models experience only a minor performance drop of less than 10\% on average when using \textit{ItemRemoved} ($S_2$) or \textit{ItemRandom} ($S_3$) instead of \textit{Original}. Although the smaller-sized model Vicuna shows a higher performance drop, it is still considerably milder compared to using \textit{ItemOnly}. To accurately reflect the recommendation abilities of LLMs with \textit{ItemRemoved} and \textit{ItemRandom}, we introduce a new post-processor denoted as $\Phi_2$ (describe in the caption of~\Cref{fig:primary}). By employing $\Phi_2$, the performance gaps between \textit{Original} and \textit{ItemRemoved} (or \textit{ItemRandom}) are further reduced. Furthermore, \Cref{fig:occurence} demonstrates the consistent and close performance gap between \textit{Original} and \textit{ItemRemoved} (or \textit{ItemRandom}) across different testing samples, which vary in size and the number of item mentions in \textit{Original}.

These results suggest that given a conversation context, LLMs primarily rely on content/context knowledge rather than collaborative knowledge to make recommendations. This behavior interestingly diverges from many traditional recommenders like collaborative filtering~\cite{koren2009mf,rendle2010fm,liang2018vaecf,he2017neural,he2018apr,sedhain2015autorec} or sequential recommenders~\cite{kang2018self,sun2019bert4rec,zhou2020s3,he2021locker}, where user-interacted items are essential.

\xhdr{Finding 5 - GPT-based LLMs possess better content/context knowledge than existing CRS.} 
From \Cref{tab:content_know}, we observe the superior recommendation performance of GPT-based LLMs against representative conversational recommendation or text-only models on all datasets, showing the remarkable zero-shot abilities in understanding user preference with the textual inputs and generating correct item titles. We conclude that GPT-based LLMs can provide more accurate recommendations than existing trained CRS models in an \textit{ItemRemoved} ($S_2$) setting, demonstrating better content/context knowledge.

\begin{table}[t]
\small
\caption{To understand the collaborative knowledge in LLMs and existing CRS models, we re-train the existing CRS models using the same perturbed conversation context \textit{ItemOnly}~($S_1$). We include the results of the representative CRS model UniCRS (denoted as CRS*) as well as a representative item-based collaborative model FISM~\cite{kabbur2013fism} (denoted as ItemCF*).}
\resizebox{\columnwidth}{!}{
\begin{tabular}{lcccccc}
\toprule
                    & \multicolumn{2}{c}{\textbf{INSPIRED}}       & \multicolumn{2}{c}{\textbf{ReDIAL}}         & \multicolumn{2}{c}{\textbf{Reddit}}         \\ \cmidrule(l){2-3} \cmidrule(l){4-5} \cmidrule(l){6-7}
\textbf{Model}     & R@1                  & R@5                  & R@1                  & R@5                  & R@1                  & R@5                  \\ \midrule
\textbf{Vicuna}             & .005 \se{.005} & .024 \se{.010} & .011 \se{.002} & .039 \se{.003} &   .005 \se{.000} &   .015 \se{.001} \\
\textbf{\texttt{GPT-3.5-t}} & .024 \se{.010} & .052 \se{.015} & .021 \se{.002} & .063 \se{.004} &   .007 \se{.001} &   .026 \se{.001} \\
\textbf{\texttt{GPT-4}}     & .014 \se{.008} & .052 \se{.015} & .025 \se{.002} & .069 \se{.004} &   \textbf{.007} \se{.001} &   \textbf{.028} \se{.001} \\ \midrule
\textbf{CRS*}               & .038 \se{.013} & .085 \se{.019} & .025 \se{.002} & .072 \se{.004} &   .003 \se{.000} &   .015 \se{.001} \\
\textbf{ItemCF*}            & \textbf{.042} \se{.012} & \textbf{.087} \se{.016} & \textbf{.029} \se{.003} & \textbf{.088} \se{.004} &   .004 \se{.001} &   .018 \se{.001} \\ \bottomrule
\end{tabular}
}
\label{tab:collaborative_know}
\end{table}

\xhdr{Finding 6 - LLMs generally possess weaker collaborative knowledge than existing CRS.} In \Cref{tab:collaborative_know}, the results from INSPIRED and ReDIAL indicate that LLMs underperform existing representative CRS or ItemCF models by 30\% when using only the item-based conversation context \textit{ItemOnly} ($S_1$). It indicates that LLMs, trained on a general corpus, typically lack the collaborative knowledge exhibited by representative models trained on the target dataset. There are several possible reasons for this weak collaborative knowledge in LLMs. First, the training corpus may not contain sufficient information for LLMs to learn the underlying item similarities. Second, although LLMs may possess some collaborative knowledge, they might not align with the interactions in the target datasets, possibly because the underlying item similarities can be highly dataset- or platform-dependent.

However, in the case of the Reddit dataset, LLMs outperform baselines in both Recall@1 and Recall@5, as shown in \Cref{tab:collaborative_know}. This outcome could be attributed to the dataset's large number of rarely interacted items, resulting in limited collaborative information. The Reddit dataset contains 12,982 items with no more than 3 mentions as responses. This poses a challenge in correctly ranking these items within the Top-5 or even Top-1 positions. LLMs, which possess at least some understanding of the semantics in item titles, have the chance to outperform baselines trained on datasets containing a large number of cold-start items.

Recent research on LLMs in traditional recommendation systems~\cite{liu2023chatgpt,hou2023large,kang2023llms} also observes the challenge of effectively leveraging collaborative information without knowing the target interaction data distribution. Additionally, another study~\cite{bao2023tallrec} on traditional recommendation systems suggests that LLMs are beneficial in a setting with many cold-start items.  Our experimental results support these findings within the context of conversational recommendations.

\begin{figure}[tbp]
  \centering
  \begin{subfigure}[b]{0.53\columnwidth}
    \includegraphics[width=\textwidth]{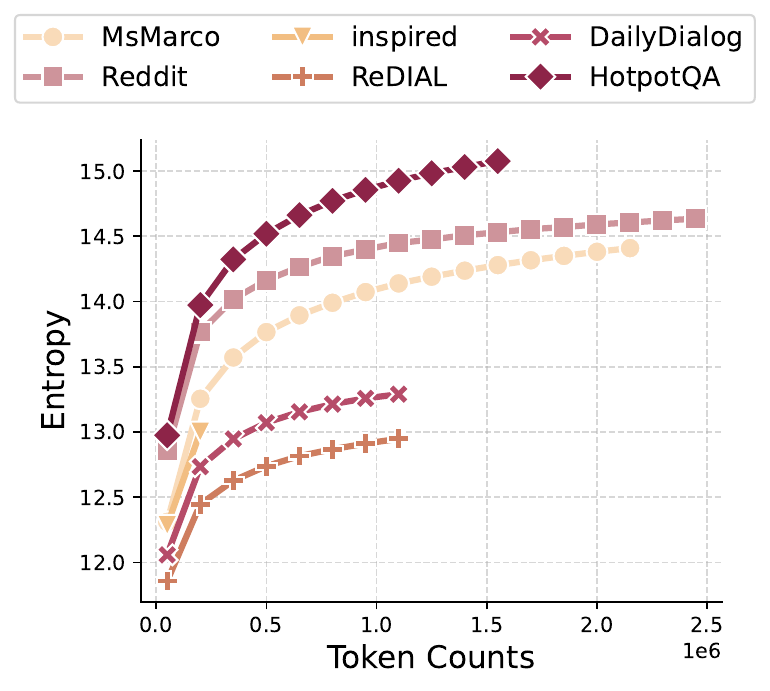}
    \caption{Entropy v.s. Token Counts}
    \label{fig:entropy}
  \end{subfigure}
  \begin{subfigure}[b]{0.43\columnwidth}
    \includegraphics[width=\textwidth]{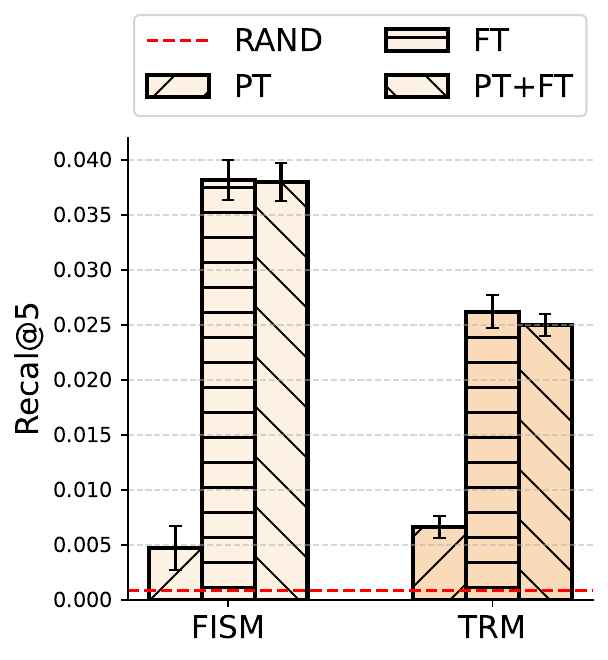}
    \caption{Pre-training Effectiveness}
    \label{fig:pre-training}
  \end{subfigure}
  \hfill
  \caption{The left subfigure shows the entropy of the frequency distribution of 1,2,3-grams with respect to number of words drawn from each dataset (item names excluded) to measure the content/context information across datasets. The right subfigure shows the results of processed Reddit collaborative dataset aligned to ML-25M~\cite{Harper2016TheMD}. RAND denotes random baseline, FT denotes fine tuning on Reddit, PT denotes pre-training on ML-25M, PT+FT means FT after PT.}
  \vspace{-8pt}
\end{figure}

\subsection{Information from CRS Data}
\label{sec:dataset-comparison}

\xhdr{Experimental Setup for Finding 7.}  
To understand LLMs in CRS tasks from the data perspective, we first measure the \textit{content/context information} in CRS datasets. Content/context information refers to the amount of information contained in conversations, excluding the item titles, which reasonably challenges existing CRS and favors LLMs according to the findings in~\Cref{sec: know_llm}.  Specifically, we conduct an entropy-based evaluation for each CRS dataset and compare the conversational datasets with several popular conversation and question-answering datasets, namely DailyDialog (chit chat)~\cite{li-etal-2017-dailydialog}, MsMarco (conversational search)~\cite{bajaj2018ms}, and HotpotQA (question answering). We use \textit{ItemRemoved} ($S_2$) conversation texts like~\Cref{sec: know_llm}, and adopt the geometric mean of the entropy distribution of 1,2,3-grams as a surrogate for the amount of information contained in the datasets, following previous work on evaluating information content in text~\cite{jhamtani2018acl}.  However, entropy naturally grows with the size of a corpus, and each CRS dataset has a different distribution of words per sentence, sentences per dialog, and corpus size. Thus, it would be unfair to compare entropy between corpus on a per-dialog, per-turn, or per-dataset basis.  To ensure a fair comparison, we repeatedly draw increasingly large subsets of texts from each of the datasets, compute the entropy of these subsets, and report the trend of entropy growth with respect to the size of the subsampled text for each CRS dataset. 

\xhdr{Finding 7 - Reddit provides more content/context information than the other two CRS datasets.}  Based on the results in~\Cref{fig:entropy}, we observe that the Reddit dataset has the most content/context information among the three conversational recommendation datasets.  Those observations are also aligned with the results in~\Cref{fig:primary,tab:content_know}, where LLMs -- which possess better content/context knowledge than baselines -- can achieve higher relative improvements compared to the other two datasets. Meanwhile, the content/context information in Reddit is close to question answering and conversational search, which is higher than existing conversational recommendation and chit-chat datasets.

\begin{figure}[tbp]
  \centering
    \includegraphics[width=0.95\columnwidth]{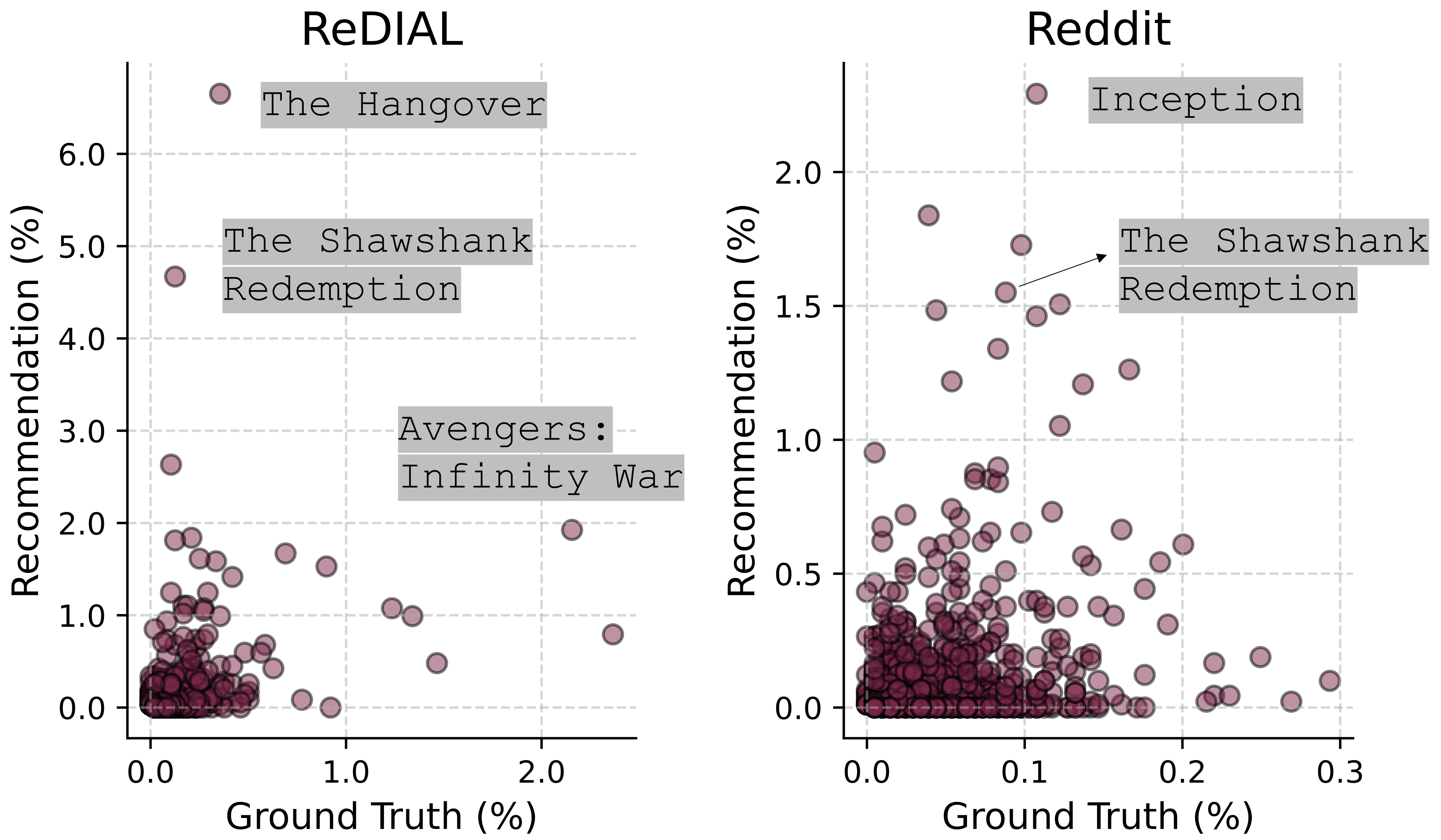}
  \caption{Scatter plots of the frequency of LLMs (\texttt{GPT-4}) generated recommendations and ground-truth items.}
  \label{fig:pop}
\end{figure}

\xhdr{Finding 8 - Collaborative information is insufficient for satisfactory recommendations, given the current models.} Quantifying the collaborative information in datasets is challenging. Instead of proposing methods to measure collaborative information, we aim to make new observations based on general performance results presented in \Cref{fig:new} and recommendation results using only collaborative information in \Cref{tab:collaborative_know}. Comparing the performance of the best models in \Cref{tab:collaborative_know} under an \textit{ItemOnly} ($S_1$) setting with the performance of the best models in \Cref{fig:new} under an \textit{Original} ($S_0$) setting reveals a significant disparity. For instance, on ReDIAL, the Recall@1 performance is 0.029 for ItemCF* compared to 0.046 for GPT-4, representing a 39.96\% decrease. Similarly, for Reddit, the Recall@1 performance is 0.007 compared to 0.023 for GPT-4 both, which is 69.57\% lower. We also experimented with other recommender systems, such as transformer-based models~\cite{kang2018self,sun2019bert4rec} to encode the item-only inputs and found similar results. Based on the current performance gap, we find that using the existing models, relying solely on collaborative information, is insufficient to provide satisfactory recommendations. We speculate that either (1) more advanced models or training methods are required to better comprehend the collaborative information in CRS datasets, or (2) the collaborative information in CRS datasets is too limited to support satisfactory recommendations.

\xhdr{Experimental Setup for Finding 9.} To understand whether the collaborative information from CRS datasets are aligned with pure interaction datasets, we conduct an experiment on the Reddit dataset. In this experiment, we first process the dataset to link the items to a popular interaction dataset ML-25M~\cite{Harper2016TheMD}~\footnote{We only use items that can be linked to ML-25M in this experiment. Here 63.32\% items are linked using the \texttt{links.csv} file from ML-25M.}. We then experiment with two representative encoders for item-based collaborative filtering based on FISM~\cite{kabbur2013fism} and Transformer~\cite{sun2019bert4rec} (TRM), respectively. We report the testing results on Reddit, with fine-tuning on Reddit (FT), pre-training on ML-25M (PT), and pre-training on ML-25M then fine-tuning Reddit (PT+FT). Note that since it is a linked dataset with additional processing, the results are not comparable with beforementioned results on Reddit.

\xhdr{Finding 9 - Collaborative information can be dataset- or platform-dependent.} From \Cref{fig:pre-training} shows that the models solely pre-trained on ML-25M (PT) outperform a random baseline, indicating that the data in CRS may share item similarities with pure interaction data from another platform to some extent. However, \Cref{fig:pre-training} also shows a notable performance gap between PT and fine-tuning on Reddit (FT). Additionally, we do not observe further performance improvement when pre-training on ML-25M then fine-tuning on Reddit (PT+FT). These observations indicate that the collaborative information and underlying item similarities, even when utilizing the same items, can be largely influenced by the specific dataset or platform. The finding also may partially explain the inferior zero-shot recommendation performance of LLMs in~\Cref{tab:collaborative_know} and suggest the necessity of further checking the alignment of collaborative knowledge in LLMs with the target datasets.

\begin{figure}[t]
  \centering
    \includegraphics[width=\columnwidth]{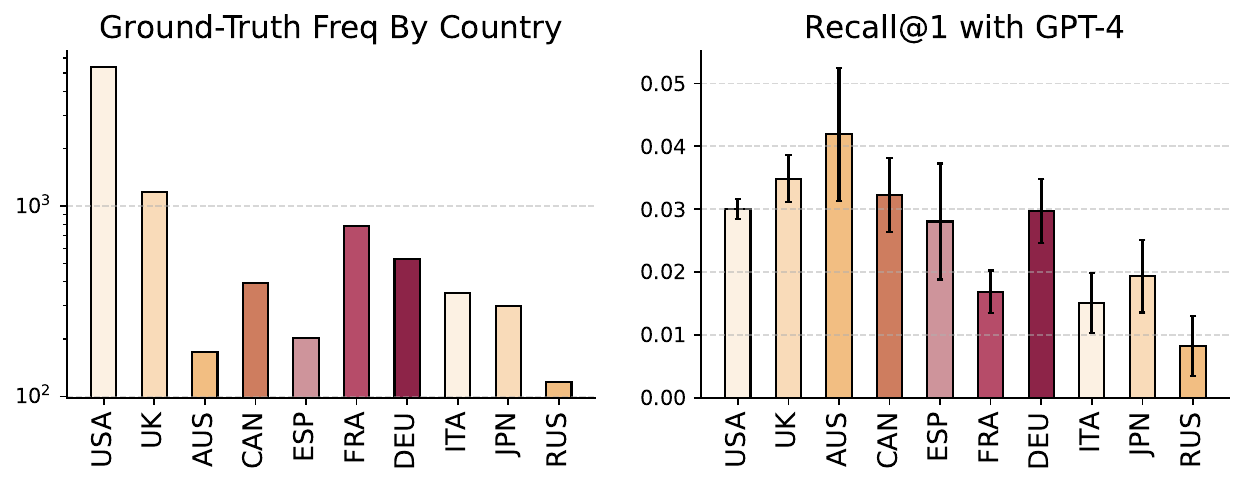}
  \caption{Ground-truth item counts in Reddit by country (in log scale) and the corresponding Recall@1 by country.}
  \label{fig:region}
\end{figure}

\subsection{Limitations of LLMs as Zero-shot CRS} 

\xhdr{Finding 10 - LLM recommendations  suffer from popularity bias in CRS.} Popularity bias refers to a phenomenon that popular items are recommended even more frequently than their popularity would warrant~\cite{chen2023bias}. \Cref{fig:pop} shows the popularity bias in LLM recommendations, though it may not be biased to the popular items in the target datasets. On ReDIAL, the most popular movies such as \texttt{Avengers: Infinity War} appear around 2\% of the time over all ground-truth items; On Reddit, the most popular movies such as \texttt{Everything Everywhere All at Once} appears less than 0.3\% of the time over ground-truth items. But for the \emph{generated} recommendations from GPT-4 (other LLMs share a similar trend), the most popular items such as \texttt{The Shawshank Redemption} appear around 5\% times on ReDIAL and around 1.5\% times on Reddit. Compared to the target datasets, LLMs recommendations are more concentrated on popular items, which may cause further issues like the bias amplification loop~\cite{chen2023bias}. Moreover, the recommended popular items are similar across different datasets, which may reflect the item popularity in the pre-training corpus of LLMs.

\xhdr{Finding 11 - Recommendation performance of LLMs is sensitive to geographical regions.} 
Despite the effectiveness in general, it is unclear whether LLMs can be good recommenders across various cultures and regions. Specifically, pre-trained language models' strong open-domain ability can be attributed to pre-training from massive data~\cite{Brown2020LMAreFewShot}. But it also leads to LLMs' sensitivity to data distribution. To investigate LLMs recommendation abilities for various regions, we take test instances from the Reddit dataset and obtain the production region of 7,476 movies from a publicly available movie dataset~\footnote{\url{https://www.kaggle.com/datasets/rounakbanik/the-movies-dataset}} by exact title matching, then report the Recall@1 for the linked movies grouped by region.
We only report regions with more than 300 data points available to ensure enough data to support the result.
As shown in ~\Cref{fig:region} the current best model, GPT-4's performance on recommendation is higher for movies produced in English-speaking regions. This could be due to bias in the training data - the left of \Cref{fig:region} show item on Reddit forums are dominated by movies from English-speaking regions.
Such a result highlights large language model's recommendation performance varies by region and culture and demonstrates the importance of cross-regional analysis and evaluation for language model-based conversational recommendation models.

 \section{Related Work}

\xhdr{Conversational Recommendation.} 
Conversational recommender systems (CRS) aim to understand user preferences and provide personalized recommendations through conversations. Typical traditional CRS setups include template-based CRS~\cite{christakopoulou2016towards,lei2020interactive,lei2020estimation, he2022bundle,zhang2022multiple} and critiquing-based CRS~\cite{chen2012critiquing,wu2019deep,li2021self}.  More recently, as natural language processing has advanced, the community developed "deep" CRS~\cite{li2018redial,chen2019kbrd,wang2022unicrs} that support interactions in natural language. 
Aside from collaborative filtering signals, prior work shows that CRS models benefit from various additional information. Examples include knowledge-enhanced models~\cite{chen2019kbrd, zhou2020kgsf} that make use of external knowledge bases~\cite{auer2007dbpedia, liu2004conceptnet}, review-aware models~\cite{lu2021revcore}, and session/sequence-based models~\cite{zou2022improving, li2022uccr}.
Presently, UniCRS~\cite{wang2022unicrs}, a model built on DialoGPT~\cite{zhang2020dialogpt} with prompt tuning~\cite{brown2020gpt3}, stands as the state-of-the-art approach on CRS datasets such as ReDIAL~\cite{li2018redial} and INSPIRED~\cite{hayati2020inspired}. Currently, by leveraging LLMs, \cite{friedman2023leveraging} proposes a new CRS pipeline but does not provide quantitative results, and \cite{wang2023rethinking} proposes better user simulators to improve evaluation strategies in LLMs. Unlike those papers,  we uncover a \textit{repeated item shortcut} in the previous evaluation protocol, and propose a framework where  LLMs serve as zero-shot CRS  with detailed analyses to support our findings from both model and data perspectives.

\xhdr{Large Language Models.} 
Advances in natural language processing (NLP) show that large language models (LLMs) exhibit strong generalization ability towards unseen tasks and domains~\cite{Chowdhery2022PaLMSL, Brown2020LMAreFewShot, Wei2022CotReasoning}. In particular, existing work reveals language models’ performance and sample efficiency on downstream tasks can be improved simply through scaling up their parameter sizes~\cite{Kaplan2020ScalingLF}.  Meanwhile, language models could further generalize to a wide range of unseen tasks by instruction tuning, learning to follow task instructions in natural language~\cite{Victor2021multitaskPromptedLm,Ouyang0JAWMZASR22TrainingLMToFollowInstruction}.
Following these advances, many works successfully deploy large language models to a wide range of downstream tasks such as question answering, numerical reasoning, code generation, and commonsense reasoning without any gradient updates~\cite{zheng2023codegeex,Brown2020LMAreFewShot,Li2022CompetitionlevelCG, Kaplan2020ScalingLF}. Recently, there have been various attempts by the recommendation community to leverage large language models for recommendation, this includes both adapting architectures used by large language models~\cite{Geng0FGZ22RecAsLp, cui2022m6rec} and repurposing existing LLMs for recommendation~\cite{li2023gpt4rec, wang2023generative, liu2023chatgpt}. 
However, to our best knowledge, we are the first work that provides a systematic quantitative analysis of LLMs' ability on \textit{conversational} recommendation.

\section{Conclusion And Discussion}

We investigate Large Language Models (LLMs) as zero-shot Conversational Recommendation Systems (CRS). Through our empirical investigation, we initially address a repetition shortcut in previous standard CRS evaluations, which can potentially lead to unreliable conclusions regarding model design. Subsequently, we demonstrate that LLMs as zero-shot CRS surpass all fine-tuned existing CRS models in our experiments. Inspired by their effectiveness, we conduct a comprehensive analysis from both the model and data perspectives to gain insights into the working mechanisms of LLMs, the characteristics of typical CRS tasks, and the limitations of using LLMs as CRS directly. Our experimental evaluations encompass two publicly available datasets, supplemented by our newly-created dataset on movie recommendations collected by scraping a popular discussion website. This dataset is the largest public CRS dataset and ensures more diverse and realistic conversations for CRS research. We also discuss the future directions based on our findings in this section.

\xhdr{On LLMs.} Given the remarkable performance even without fine-tuning, LLMs hold great promise as an effective approach for CRS tasks by offering superior content/contextual knowledge. The encouraging performance from the open-sourced LLMs~\cite{xu2023baize, vicuna2023} also opens up the opportunities to further improve CRS performance via efficient tuning~\cite{hu2021lora, bao2023tallrec} and collaborative filtering~\cite{koren2009mf} ensembling. Meanwhile, many conventional tasks, such as debiasing~\cite{chen2023bias} and trustworthy~\cite{ge2022survey} need to be revisited in the context of LLMs.

\xhdr{On CRS.} Our findings suggest the systematic re-benchmarking of more CRS models to understand their recommendation abilities and the characteristics of CRS tasks comprehensively. Gaining a deeper understanding of CRS tasks also requires new datasets from diverse sources e.g.,~crowd-sourcing platforms~\cite{li2018redial, hayati2020inspired}, discussion forums, and realistic CRS applications with various domains, languages, and cultures. Meanwhile, our analysis of the information types uncovers the unique importance of the superior content/context knowledge in LLMs for CRS tasks; this distinction also sets CRS tasks apart from traditional recommendation settings and urges us to explore the interconnections between CRS tasks and traditional \textit{recommendation}~\cite{Harper2016TheMD} or \textit{conversational search}~\cite{bajaj2018ms} tasks.

\newpage

\bibliographystyle{ACM-Reference-Format}
\bibliography{main}

\end{document}